\documentclass[preprint2]{aastex}
\usepackage{color}
\usepackage{amsmath}
\usepackage{natbib} 

\slugcomment{To appear in PASP}
\shorttitle{Measurements and Mitigation of Reciprocity Failure}
\shortauthors{Biesiadzinski \textit{et al.}}

\begin{document}

\title{Reciprocity Failure in HgCdTe
Detectors: Measurements and Mitigation}

\author{
T.~Biesiadzinski,
W.~Lorenzon, R.~Newman,
M.~Schubnell, G.~Tarl\'e,
C.~Weaverdyck}

\begin{abstract}

A detailed study of reciprocity failure in four 1.7\,$\mu$m cutoff HgCdTe
near-infrared detectors is presented. The sensitivity to reciprocity failure is
approximately 0.1\,\%\,decade$^{-1}$ over up to five orders of magnitude in
illumination intensity. The four detectors, which represent three successive
production runs with modified growth recipes, show large differences in amount
and spatial structure of reciprocity failure. Reciprocity failure could be
reduced to negligible levels by cooling the detectors to about 110\,K. No
wavelength dependence was observed. The observed spatial structure appears to
be weakly correlated with image persistence.

\end{abstract}

\affil{Department of Physics, University of Michigan, Ann Arbor, MI 48109}


\maketitle

\section{Introduction}

Nonlinearity in detector response can severely impact photometric precision in
astronomical observations. Recent reports of count rate dependent nonlinearity
observed in HgCdTe near-infrared (NIR) detectors \citep{boh2005, rie2010,
hil2010, deu2010, bie2011} suggest that this effect is common in HgCdTe
detectors, although so far only measurements with detectors from the
HAWAII\footnote{HAWAII: HgCdTe Astronomy Wide Area Infrared Imager} family
produced by Rockwell Science Center, now Teledyne Imaging Sensors (TIS), have
been reported. Count rate dependent nonlinearity is also referred to as flux
dependent nonlinearity or as reciprocity failure. It describes the failure of a
detector to respond linearly to changes in the incident flux. A power law of
the form $count rate \propto flux^{\,\alpha}$ can be used to model the effect,
where $\alpha$ represents the amount of nonlinearity. Values for nonlinearity
from less than  0.1\,\%\,decade$^{-1}$ to about 10\,\%\,decade$^{-1}$ have been
reported. Reciprocity failure was first pointed out in HgCdTe detectors by
\cite{boh2005} for the 2.5\,$\mu$m cutoff Near Infrared Camera and Multi-Object
Spectrometer (NICMOS) detectors. For the NICMOS detectors a nonlinearity of
about 6\,\%\,decade$^{-1}$ was reported based on a comparison of NICMOS and
STIS (Space Telescope Imaging Spectrograph) standard star observations
\citep{boh2005}. Further evidence for reciprocity failure comes from
measurements on 1.7\,$\mu$m cutoff detectors produced for the NIR channel of
the Wide Field Camera 3 (WFC3), installed onboard the Hubble Space Telescope
during the final servicing mission in 2009. These detectors show reciprocity
failure between 0.3\,\% and 1\,\% in the wavelength range from 0.85\,$\mu$m to
1.0\,$\mu$m \citep{hil2010}.

In all measured detectors this nonlinearity is such that for a constant
fluence, a pixel's response to a high flux is larger than its response to a low
flux. This nonlinearity is different from the well-known and well-understood
classical nonlinearity that is observed as  charge is integrated at the
junction capacitance of the pixel node. The physical mechanism that leads to
reciprocity failure however is not yet understood. Charge traps in the bulk
material have been suggested as the cause for image persistence
\citep{smith08}, and it is conceivable that a mechanism based on charge
trapping is also responsible for reciprocity failure. Alternatively, this
nonlinearity could originate in the HAWAII multiplexer, or it may be caused by
small leakage currents at the charge integrating transistors.

\section{Instrument and Operation}

A dedicated test system was designed and built to precisely characterize
reciprocity failure in HgCdTe detectors over their full spectral and dynamic
range. A sensitivity to reciprocity failure of approximately 0.1\,\%  was
achieved for flux levels between 1 and 50,000 photons\,s$^{-1}$ and exposure
times of 5 seconds to 5 hours. A detailed description of this setup can be
found in \cite{bie2011}. In the following only a brief outline of the
experimental method is presented followed by a discussion of experimental
uncertainty in the linearity measurement and the readout strategy.

To quantify reciprocity failure, the HgCdTe detector response was measured and
compared to the response of a monitoring photodiode while the illumination
intensity was varied. The detector was repeatedly read out in what is referred
to as sample-up-the-ramp mode (SUR). The analysis of the detector response at
different charge integration levels allows for the removal of the classical
gain non-linearity \citep{bie2011}. For this measurement the detector dark
current, typically on the order of 0.05 e/sec/pixel, was taken into
account.\footnote{Any background illumination, including thermal leakage, was
smaller than the dark current level and was treated as a part of the dark
current.}

In order to obtain an absolute measurement of reciprocity failure, knowledge of
the linearity of the monitoring photodiodes over the full illumination level
used during the investigation was required. Good spectral coverage was achieved
by using two monitoring diodes, a Si photodiode for measurements up to
1.0\,$\mu$m and an InGaAs photodiode for measurements up to 1.8\,$\mu$m.
Precise linearity characterization of the diodes was not available from the
vendor. Thus linearity of the Si photodiode (Edmund Optics 53371) was measured
independently in a dedicated setup \citep{bie2011}. The Si photodiode
illumination intensity nonlinearity was measured to be
(0.08\,$\pm$\,0.08)\,\%\,decade$^{-1}$ consistent with  zero. Along with the
comparison of Si and InGaAs photodiode responses, this established a
0.1\%\,decade$^{-1}$ systematic uncertainty in our test setup.

The reciprocity characterization setup operates shutter-less and therefore the
shortest exposure time is determined by the time it takes to read out the
detector. This limits the dynamic range of observable photon intensities. To
extend this range towards higher photon fluxes the majority of the
characterization was performed in a mode in which only part of the detector was
read out. This readout mode is referred to as \emph{stripe mode}. In this mode
only an area of 300\,$\times$\,2048 pixels was read out to reduce the readout
time from the 1.418 seconds it takes to read the full detector
(2048\,$\times$\,2048 pixels) to 0.210 seconds.

A fraction of the measurements was performed where the full detector was read
out to probe possible spatial variation in reciprocity failure across a
detector. This readout mode is referred to as \emph{full mode}. The spatial
resolution was sampled by subdividing the detector into tiles of
64\,$\times$\,64 pixels in the \emph{full mode} and 60\,$\times$\,64 pixels in
the \emph{stripe mode}. This tiling reduces the uncertainty in the measurement
due to photon shot noise and read noise.

\section{Measurements and Results}

The SuperNova\,/\,Acceleration Probe (SNAP) was proposed as a satellite mission
to explore the nature of dark energy \citep{ald2002}. SNAP evolved into JDEM,
the Joint Dark Energy Mission. As part of the R\&D effort for SNAP/\,JDEM a
number of near-infrared detectors were procured and characterized.  The SNAP
science specifications, challenging vendor capabilities at the onset of the
program, could ultimately be met and several devices with low read noise and
very good quantum efficiency (QE) were delivered \citep{schub06}. In the course
of several production runs the HgCdTe growth and detector manufacturing
processes were tuned to improve performance. The detector growth recipe was
originally largely based on the experience gained with devices produced for
WFC3. The WFC3 baseline process was extended to a 2048\,$\times$\,2048 format
with an emphasis on reducing read noise while maintaining good quantum
efficiency. This development ran parallel to the SNAP effort and both projects
aimed to improve detector quantum efficiency and achieve low read noise and
dark current. Initially, the improvement of those detector characteristics was
driving development efforts, but later fabrication runs also included substrate
removal and reduction of capacitive coupling between pixels. For the
investigation presented here, four SNAP/\,JDEM devices, representing three
generations of development, were tested in the University of Michigan
reciprocity characterization system. All four detectors were produced by TIS
and all have 2048\,$\times$\,2048 pixels with a pixel pitch of 18\,$\mu$m. They
are HgCdTe detectors with a high wavelength cutoff at 1.7\,$\mu$m hybridized to
a \mbox{HAWAII-2} readout integrated circuit. Substrate removal and
anti-reflective coating provide good response extending from the UV to the
near-infrared. Detector characterization was initially performed at a single
temperature (140 K) followed by measurements at several wavelengths to test a
possible wavelength dependence of reciprocity failure. In later measurements
the temperature was also varied to investigate temperature dependence. In
section \ref{sec:measurements_140} the results from measurements on each of the
four detectors at 140\,K are discussed. In section \ref{sec:temperature} it is
shown how reciprocity failure can be mitigated by lowering the device
temperature. An overview of all measurements can be found in Tables
\ref{tab:recip_table} and \ref{tab:temperature_table}. Note that the quoted
uncertainties are statistical only and do not include the overall 0.1\%
systematic uncertainty.

\subsection{Measurements at 140\,K}
\label{sec:measurements_140}

\subsubsection{H2RG-102}

Device H2RG-102 was manufactured early on during the SNAP/JDEM R\&D program and
was delivered in 2005. The QE is greater than 90\% from 0.9\,$\mu$m to
1.7\,$\mu$m  and about 40\% at 0.45\,$\mu$m. The dark current and read noise
performance is very good; the Fowler-1 noise is 25 e$^-$. Unlike devices
produced later, this detector is mounted on a molybdenum pedestal. The
multiplexer is of type HAWAII-2RG-A0.

Characterization of reciprocity failure in detector H2RG-102 was described in
detail in \citet{bie2011}. This detector exhibits low reciprocity failure
($0.35\,\pm\,0.03$)\,\%\,decade$^{-1}$ at 790\,nm and shows no wavelength
dependence.

\subsubsection{H2RG-142}

\begin{figure}[!htbp]
\begin{center}
\includegraphics[width=0.98\linewidth]{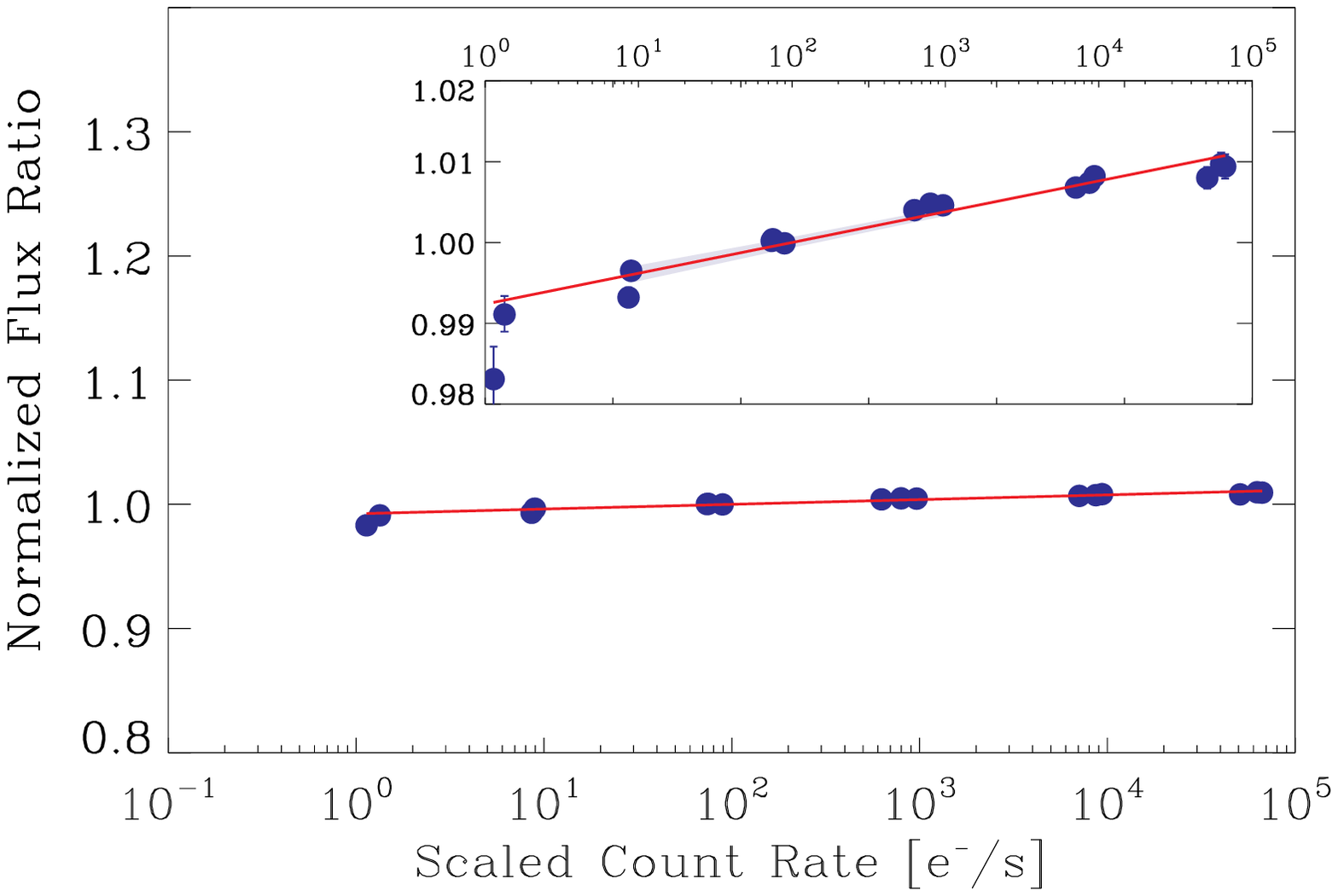}
\caption{Reciprocity failure measured in \emph{stripe mode}
	for device H2RG-142 at 790 nm. The ordinate scale was set to
	allow a direct comparison with other
	detectors. A magnified scale is shown in the insert.
	The 68\,\% confidence level is indicated by the shaded area.}
\label{fig:dev_142_strip}
\end{center}
\end{figure}

Device H2RG-142 came from the fifth manufacturing run for SNAP. It was
mounted on a SiC pedestal specifically developed for SNAP/JDEM to provide a
good thermal match to the multiplexer. Devices from this run were also mated
to the HAWAII-2RG-A0 multiplexer. H2RG-142 has high QE and low read noise. It
exhibits a somewhat larger number of hot pixels than H2RG-102 but is
otherwise cosmetically good. Figure \ref{fig:dev_142_strip} shows reciprocity
failure of \mbox{($0.38\ \pm\ 0.03$)\,\%\,decade$^{-1}$} at 790\,nm in
\emph{stripe mode}. The average reciprocity failure value measured for this
device was very similar to detector H2RG-102 at all wavelengths.

In addition to the \emph{stripe mode} measurements the structure of reciprocity
failure was also characterized in the \emph{full mode}. Although the signal to
noise ratio was low, nonlinearity variations in the detector did appear in a
range from 0.35 to 0.85\,\%\,decade$^{-1}$. In particular one corner of the
device exhibited larger reciprocity failure.

\subsubsection{H2RG-236 and H2RG-238}

\begin{figure}[!htbp]
\begin{center}
\includegraphics[width=0.98\linewidth]{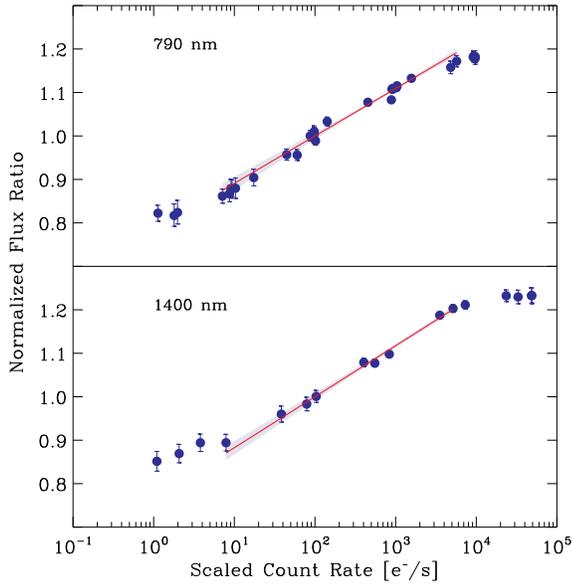}
\caption{Average reciprocity failure measured in device H2RG-236
    at 790\,nm (top panel) and 1400\,nm (bottom panel). Data was taken in the \emph{stripe mode}.
    The 68\,\% confidence
	level is indicated by the shaded area. }
\label{fig:dev_236_strip}
\end{center}
\end{figure}

Devices H2RG-236 and H2RG-238 were produced during the sixth manufacturing run
of the SNAP\,/JDEM R\&D program. Like device H2RG-142 they both are mounted on
a SiC pedestal but unlike that device, they were hybridized to a newer
multiplexer, the HAWAII-2RG-A1 designed in part to reduce capacitive coupling
between neighboring pixels \citep{bro2006}. Both devices have low dark current
and read noise and are very good cosmetically. Quantum efficiency of both
devices is lower than in earlier detectors but is exceptionally uniform when
measured at high flux. The average reciprocity failure measured in \emph{stripe
mode} for device H2RG-236, shown in Figure \ref{fig:dev_236_strip}, is
($10.9\,\pm\,0.5$)\,\%\,decade$^{-1}$ and ($11.7\,\pm\,0.5$)\,\%\,decade$^{-1}$
at 790\,nm and 1400\,nm, respectively. The results from the two measurements
are very similar, emphasizing the insensitivity of reciprocity failure to the
wavelength of the illumination for these detectors. Data taken at 1400\,nm and
950\,nm (not shown in Figure~\ref{fig:dev_236_strip}) revealed that a linear
fit is only representative for illumination levels between roughly 10
counts\,s$^{-1}$ and 10,000 counts\,s$^{-1}$. Outside this range the detector
response appears to become linear, indicating a saturation effect at high
illumination levels and possibly a turn-on threshold at low illumination
levels. Detector H2RG-236 showed the largest reciprocity failure of the four
devices measured.

\begin{figure}[!htbp]
\begin{center}
\includegraphics[width=0.98\linewidth]{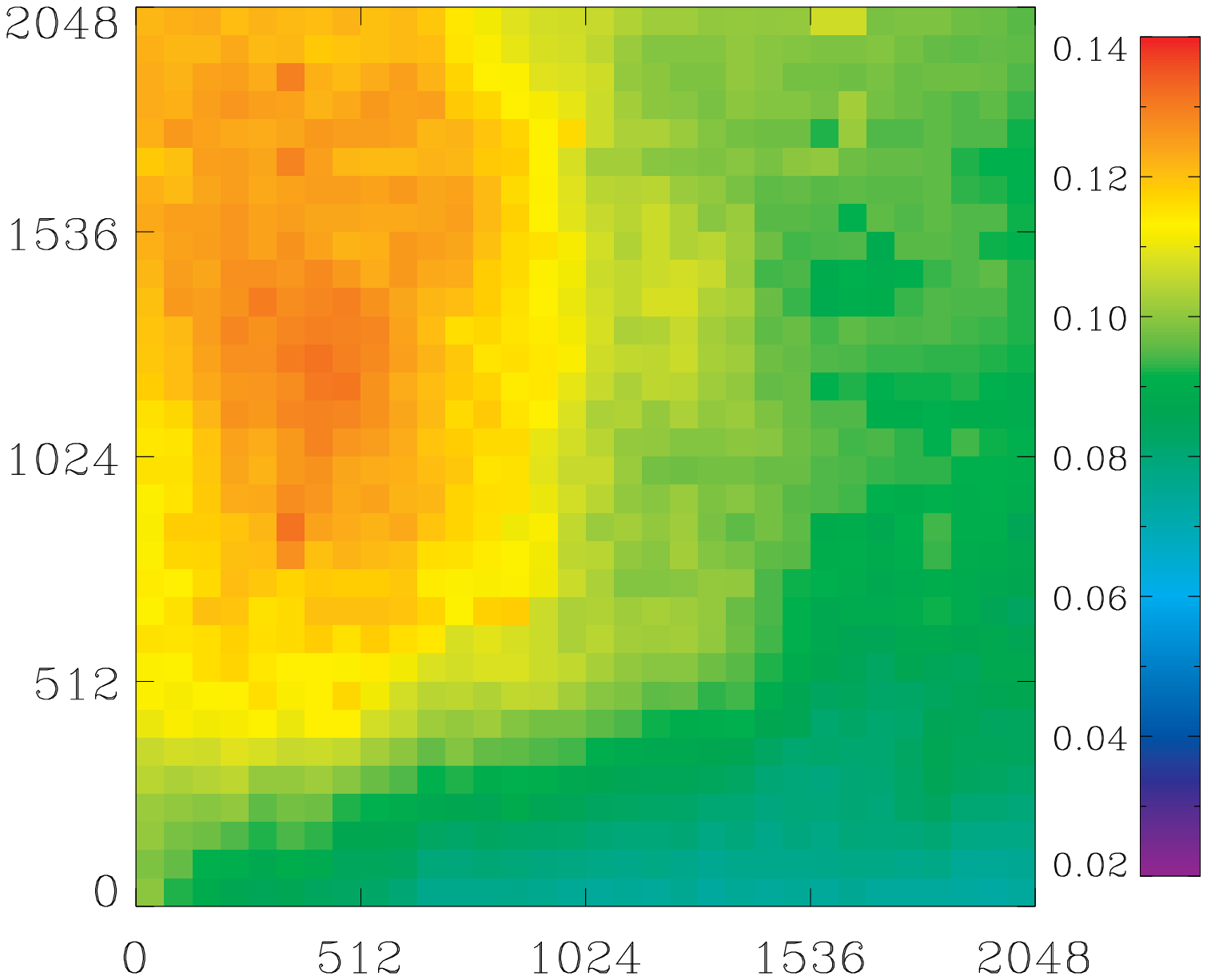}
\caption{Reciprocity failure map for device H2RG-236 at 790 nm. The scale is in \%\,decade$^{-1}$.}
\label{fig:dev_236_full}
\end{center}
\end{figure}

\begin{figure}[!htbp]
\begin{center}
\includegraphics[width=0.98\linewidth]{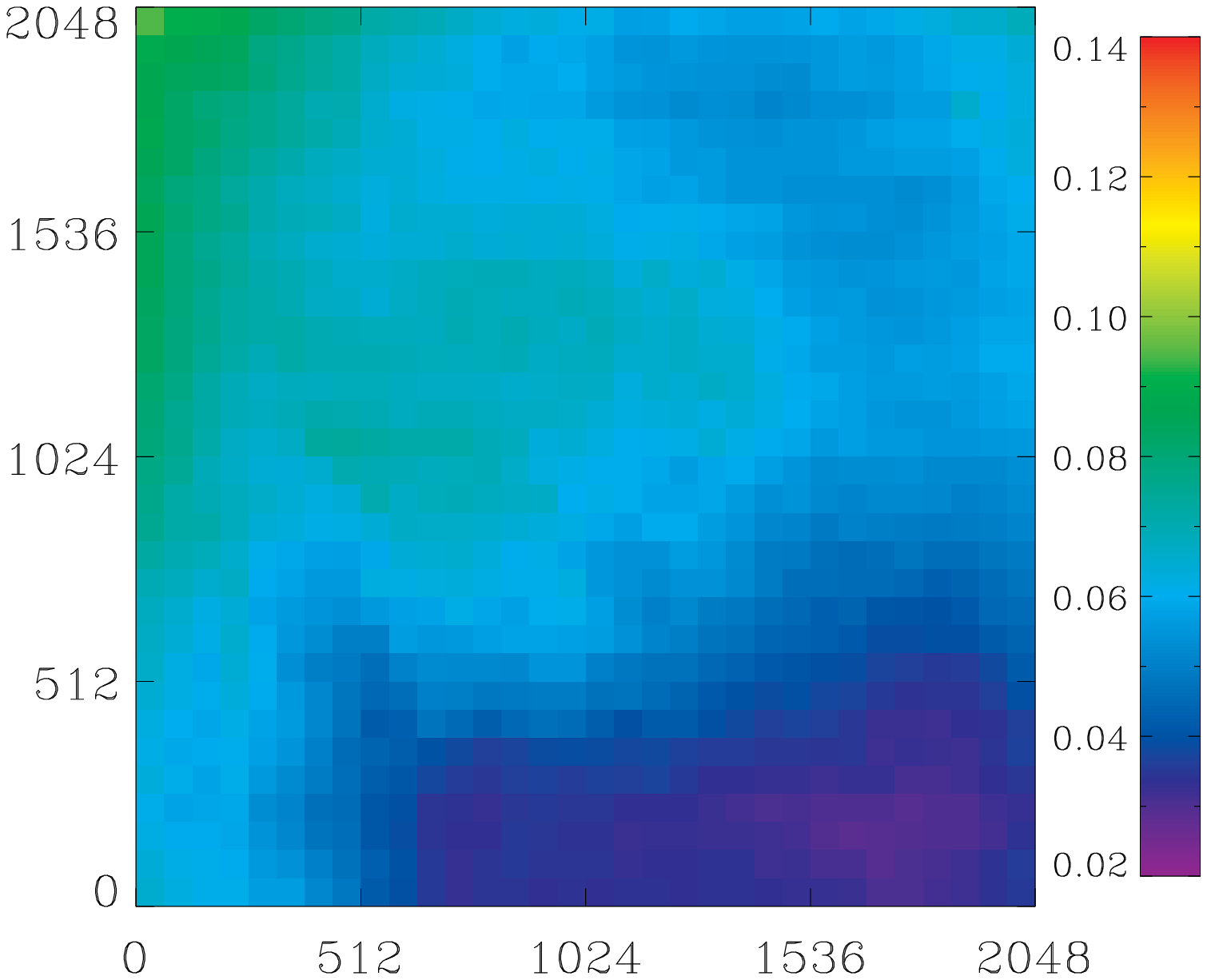}
\caption{Reciprocity failure map for device H2RG-238 at 950 nm. The scale is in \%\,decade$^{-1}$.}
\label{fig:dev_238_full}
\end{center}
\end{figure}

Strong spatial variation of reciprocity failure was observed in these two
devices, ranging from 7.3\,\%\,decade$^{-1}$ to 13.1\,\%\,decade$^{-1}$ for
device H2RG-236 (Figure \ref{fig:dev_236_full}) and from 2.9\,\%\,decade$^{-1}$
to 9.5\,\%\,decade$^{-1}$ for device H2RG-238 (Figure \ref{fig:dev_238_full}).
It is worth noting that for such a device, simply correcting for the average
reciprocity failure without accounting for spatial structure will result in a
large residual uncertainty in photometric measurements.

\subsection{Temperature Dependence}
\label{sec:temperature}

\begin{figure}[!htbp]
\begin{center}
\includegraphics[width=0.94\linewidth]{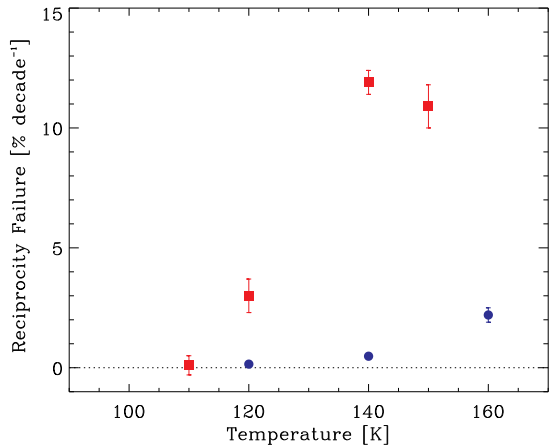}
\caption{Reciprocity failure as a function of detector
    temperature for devices H2RG-142 (blue circles) and H2RG-236 (red squares).}
\label{fig:temp}
\end{center}
\end{figure}

In an attempt to better understand the physical mechanisms that lead to
reciprocity failure, it was investigated how reciprocity failure is affected by
device temperature. Detectors H2RG-142 and H2RG-236, low and high reciprocity
devices, respectively, were tested at temperatures ranging from 100\,K to
160\,K.  These tests revealed that flux dependent nonlinearity can be ``frozen
out'' at sufficiently low temperatures. The results from the two detectors,
shown in Figure \ref{fig:temp}, suggest that this freeze-out temperature
depends on the amount of reciprocity failure in a particular detector and will
therefore most likely vary for different detectors. An overview of the
temperature test results is presented in Table \ref{tab:temperature_table}.

\subsection{Reciprocity Failure and QE}

For detectors that exhibit reciprocity failure, care must be taken when
measuring quantum efficiency. Reciprocity failure will bias QE measurements
towards higher values at high illumination levels and towards lower QE values
at low illumination levels. In addition, spatial nonuniformity of reciprocity
failure across a detector will alter the apparent device uniformity as a
function of the illumination intensity. One possible approach is to measure
QE at sufficiently low temperature to suppress reciprocity failure in order
to reveal the ``true'' QE.

\begin{figure}[!htbp]
\begin{center}
\includegraphics[width=0.94\linewidth]{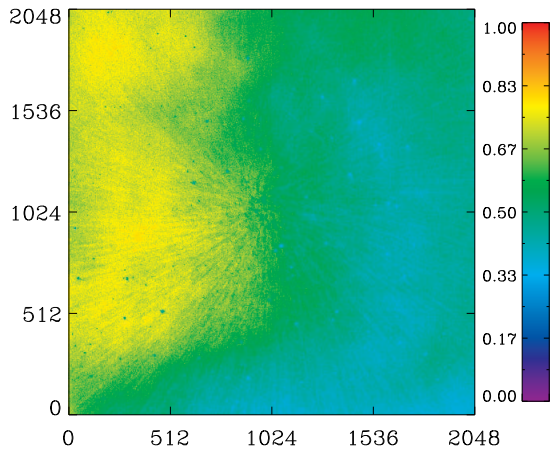}
\caption{Ratio of two H2RG-236 flat field images with a
    factor of 1000 difference in flux. The observed large scale structure
    is due to reciprocity
    failure. The measurement was performed at 140\,K. }
\label{fig:recip_and_qe236}
\end{center}
\end{figure}

Precise characterization of reciprocity failure is a rather elaborate
procedure and requires a specialized experimental setup. However, a simple
measurement can reveal possible \emph{spatial structure} in a detector's
reciprocity failure. Using a standard flat field illumination test setup two
flat field images were produced, one at a very high illumination intensity
and a second at a very low illumination intensity. The ratio of these two
images, shown in Figure \ref{fig:recip_and_qe236}, displays the same spatial
variability as the reciprocity failure map for this detector shown in Figure
\ref{fig:dev_236_full}. Such a measurement may therefore be used as a simple
test that does not require any special equipment beyond a basic illumination
system. However, some caveats apply. This test will only reveal spatial
structure in reciprocity failure of a device, and will not produce an
absolute value for the strength of reciprocity failure, nor will it reveal
reciprocity failure in detectors where the effect is spatially uniform.

\section{Discussion}

Although a detector's reciprocity failure can be large, it will likely be
possible to correct for it. If a sufficient amount of calibration data is
obtained it should be possible to correct for reciprocity failure on a pixel by
pixel level. Cooling detectors that exhibit strong reciprocity failure provides
a straightforward mitigation strategy although the required temperature may
vary for individual devices. While, for example, detector H2RG-142 will likely
not exhibit noticeable reciprocity failure at 120\,K, device H2RG-236 would
have to be cooled below 100\,K.

The observed spatial nonuniformity in reciprocity failure provides an
opportunity to investigate a possible correlation with other detector
properties such as dark current, QE near cutoff\,\footnote{In all HgCdTe
devices that were tested, strong QE variations are observed near the cutoff
wavelength. This is caused by inconsistencies in the doping of the HgCdTe
material by the MBE process.}, and image persistence. Therefore the cross
correlation between the spatial structure of reciprocity failure and the other
properties was computed for device H2RG-236. This particular detector was
selected because of the pronounced spatial nonuniformity in reciprocity
failure. The correlation coefficient was normalized to have a value between
$-1$ and $1$ for fully anti-correlated structure and identical structure,
respectively. A value of zero represents the absence of correlation. Both, the
790\,nm  and the 1400\,nm reciprocity failure data were used for this analysis
as shown in Table \ref{tab:pattern_corr}. The correlation coefficient for the
reciprocity failure maps at those two wavelengths is 0.92, indicating that not
only the average reciprocity is independent of wavelength but also the spatial
structure.

The two detectors that show low reciprocity failure, H2RG-102 and H2RG-142, and
the two detectors that show high reciprocity failure, H2RG-236 and H2RG-238,
differ in the type of multiplexer used for device readout. The 100-series
detectors were hybridized to the HAWAII-2RG-A0 multiplexer while for the
200-series the redesigned HAWAII-2RG-A1 multiplexer was used. It was
investigated whether the change in the multiplexer design was responsible for
the large discrepancy in reciprocity failure between the 100 and 200 series.
For this test an external RC circuit with a large capacitance and a precisely
measured selectable resistance was used. The RC circuit was charged, simulating
charge collecting at the pixel node, and read out by the multiplexer. Using the
RC circuit instead of the detector diode allowed to measure the linearity
response of the multiplexer by varying the circuit's impedance. The test was
performed with the multiplexers of devices H2RG-142 and H2RG-238. No difference
in multiplexer voltage readout linearity was observed, indicating that the
difference in multiplexer readout electronics alone is not responsible for the
observed difference in reciprocity failure.

At present the fundamental mechanism that leads to reciprocity failure is not
understood. The comparison of spatial structures in characteristic maps
discussed above does not provide a satisfactory suggestion of correlation
between reciprocity failure and any other detector characteristic. In fact the
only correlation that has been observed so far is that detectors with high
reciprocity failure show also large image persistence, and detectors that show
very low reciprocity failure tend to have low image persistence. Note that this
observation is based on the very limited sample of detectors discussed here and
may not be a general property of HgCdTe detectors. Small leakage currents due
to Ohmic parasitic resistance across the integrating field effect transistor
can be excluded as cause for reciprocity failure because they would not
reproduce the observed power-law behavior. However, non-linear leakage
currents, typical for diodes, may provide an explanation for this effect.
Furthermore, a charge trapping mechanism has been suggested as the underlying
mechanism for image persistence \citep{smith08}, and it is conceivable that
such a process also accounts for reciprocity failure.

\section{Conclusions}
Reciprocity failure was measured in four devices developed as part of the
SNAP/JDEM R\&D program with an overall sensitivity of 0.1\% per decade in
illumination intensity. It was found to vary from device to device with
detector-averaged values (in \%\,decade$^{-1}$ at 790\,nm) of $0.35\,\pm\,0.03$
for H2RG-102, $0.38\,\pm\,0.03$ for H2RG-142, $10.9\,\pm\,0.5$ for H2RG-236 and
$5.1\ \pm\ 0.7$ for H2RG-236. In addition, spatial variation of reciprocity
failure was observed in all three devices that were tested in the \emph{full}
readout mode. A wavelength dependence, such as reported for the NICMOS
detectors, was not observed.

Reciprocity failure causes a systematic error in measurements of faint
astronomical sources relative to bright standards. If not corrected for, an
observation spanning three decades in illumination could suffer from a 1\% (in
low reciprocity devices) to 30\% (in high reciprocity devices) error in the
flux determination. Such a device would, if used for supernova cosmology for
example, lead to an incorrect overestimate of the acceleration of the universe.
In addition, this nonlinearity has to be accounted for when performing a
standard detector characterization such as measuring QE. The value of QE and
its spatial uniformity depends on the intensity of the light at which they are
measured.

Because of the wide range of reciprocity failure from one detector to another
and of its spatial structure, reciprocity failure calibration presents a
challenge. Furthermore, it is currently unknown if on-orbit radiation damage
may alter it. Without a fundamental understanding of the underlying
mechanism, reciprocity failure is therefore best addressed by the selection
of ``low reciprocity failure'' devices and by cooling them sufficiently.

\acknowledgments We gratefully acknowledge the many
valuable conversations with Roger Smith and Christopher
Bebek during the course of this work. This work was
supported by the Director, Office of Science, of the U.S.
Department of Energy under Contract Nos.
DE-FG02-95ER40899 and DE-FG02-08ER41566.\\[30mm]

\clearpage

\begin{figure}[!htbp]
\begin{center}
\includegraphics[width=0.94\linewidth]{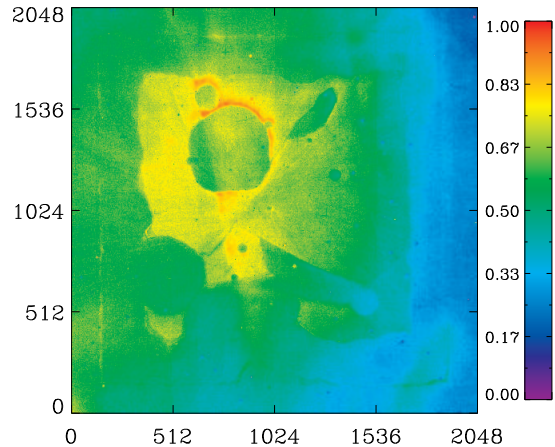}
\caption{Image persistence in H2RG-236. }
\label{fig:pers236_full}
\end{center}
\end{figure}

\clearpage

\begin{table}[!htbp]
\begin{center}
\caption{Reciprocity failure data at 140\,K.\label{tab:recip_table}}
\begin{tabular}{ lccc }
\tableline\tableline
				& Wave-		& \multicolumn{2}{c}{Reciprocity Failure} \\
	Detector		& length		& \emph{Stripe Mode} & \emph{Full Mode}  \\
				& [nm]		& \multicolumn{2}{c}{[\%\,decade$^{-1}$]} \\
\tableline
	H2RG-102\tablenotemark{a} & 700	& 0.35 $\pm$ 0.04	& \nodata\\
				& 790		& 0.35 $\pm$ 0.03	& \nodata \\
				& 880		& 0.36 $\pm$ 0.05	&\nodata \\
				& 950		& 0.29 $\pm$ 0.04	& \nodata\\
				& 1400		& 0.38 $\pm$ 0.05	& \nodata\\
	\\
	H2RG-142	& 790		& 0.38 $\pm$ 0.03	& 0.53 $\pm$ 0.14 \\
				& 950		&  0.48 $\pm$ 0.07	& \nodata\\
				& 1400		& 0.33 $\pm$ 0.04	& \nodata\\
	\\
	H2RG-236	& 790		& 10.9 $\pm$ 0.5	& 10.3 $\pm$ 0.6 \\
				& 950		& 11.9 $\pm$ 0.5	& \nodata\\
				& 1400		& 11.7 $\pm$ 0.5	& 10.6 $\pm$ 1.9\\
	\\
	H2RG-238 	& 790 		& 5.1 $\pm$ 0.7\tablenotemark{b}	& 4.0 $\pm$ 0.8 \\
				& 950 		& 4.4 $\pm$ 0.4 	& \nodata\\	
\tableline
\end{tabular}
\tablenotetext{a}{Published in \citet{bie2011}.}
\tablenotetext{b}{ \emph{Full mode} data analyzed as \emph{stripe mode}.}
\end{center}
\end{table}

\begin{table}[!htbp]
\begin{center}
\caption{Reciprocity failure versus temperature. \label{tab:temperature_table}}
\begin{tabular}{ c c c}
\\
	\tableline
		Temperature	& \multicolumn{2}{c}{Reciprocity Failure} \\
		{[K]}		& \multicolumn{2}{c}{[\%\,decade$^{-1}$]} \\		
	  			& H2RG-142				& H2RG-236 \\
	\tableline\tableline
		160		& 2.2 $\pm$ 0.3			& \nodata \\
		150		& 	\nodata					& 10.9 $\pm$ 0.9 \\
		140		& 0.48 $\pm$ 0.07 		& 11.9 $\pm$ 0.5 \\
		120		& 0.15 $\pm$ 0.07 		& 3.0 $\pm$ 0.7 \\
		100		& 	\nodata					& 0.1 $\pm$ 0.4 \\
		\tableline
\end{tabular}
\tablecomments{Data was obtained in \emph{stripe mode} at 950nm.}
\end{center}
\end{table}

\begin{table}[!htbp]
\begin{center}
\caption{Correlation of reciprocity failure and other detector properties at
790\,nm and 1400\,nm. \label{tab:pattern_corr}}
\begin{tabular}{lcc}
\\
	\tableline
			& 790\,nm	& 1400\,nm				\\
				
	\tableline\tableline

				&		&					\\
	Dark Current	& -0.41	& -0.42			\\
	QE at 1750 nm	& 0.11	& 0.15	 	\\
	Persistence		& 0.70	& 0.57	 	\\
	Conversion Gain			& -0.09	& 0.00		\\
		&		&			\\
	\tableline
\end{tabular}
\end{center}
\end{table}

\end{document}